\documentclass[prd, showpacs, showkeys, twocolumn]{revtex4}
\usepackage{bm}
\usepackage{amssymb,amsmath,amsthm}
\usepackage{mathrsfs}
\begin{document}
\title{Quantum Interest in (3+1) dimensional Minkowski space}
\author{Gabriel Abreu}
\email{gabriel.abreu@mcs.vuw.ac.nz}
\affiliation{School of Mathematics, Statistics, and Computer Science, 
Victoria University of Wellington, PO Box 600, Wellington, New Zealand\\}
\author{Matt Visser}
\email{matt.visser@mcs.vuw.ac.nz}
\affiliation{School of Mathematics, Statistics, and Computer Science, 
Victoria University of Wellington, PO Box 600, Wellington, New Zealand\\}

\date{14 August 2008; Revised 8 January 2009; 
\LaTeX-ed \today}
\begin{abstract}

The so-called ``Quantum Inequalities'', and the ``Quantum Interest Conjecture'', use quantum field theory to impose significant restrictions on the temporal distribution of the energy density measured by a time-like observer, potentially preventing the existence of exotic phenomena such as ``Alcubierre warp-drives'' or ``traversable wormholes''. Both the quantum inequalities and the quantum interest conjecture can be reduced to statements concerning the existence or non-existence of bound states for a certain one-dimensional quantum mechanical pseudo-Hamiltonian. Using this approach, we shall provide a simple variational proof of one version of the Quantum Interest Conjecture in (3+1) dimensional Minkowski space.

\end{abstract}


\maketitle

\def\d{{\mathrm{d}}}
\def\implies{\Rightarrow}
\newcommand{\scri}{\mathscr{I}}
\newcommand{\sun}{\ensuremath{\odot}}
\def\tr{{\mathrm{tr}}}
\def\sech{{\mathrm{sech}}}
\def\etc{\emph{etc}}
\def\ie{{\emph{i.e.}}}
\newtheorem{theorem}{Theorem}
\newtheorem{corollary}{Corollary}
\def\lint{\hbox{\Large $\displaystyle\int$}} 
\def\hint{\hbox{\Huge $\displaystyle\int$}}  
\def\Re{{\mathrm{Re}}}
\def\Im{{\mathrm{Im}}}

\section{Introduction}

It is well-known that quantum physics permits arbitrarily large negative energy densities at individual points \cite{epstein-glaser-jaffe, roman-wec}, though averages and total energies over volumes or  lines are much more tightly constrained. This is of critical importance when developing singularity theorems, and other theorems based on global analysis, in that this simple observation is enough to guarantee that the so-called ``classical energy conditions'' are not fundamental physics; they are at best classical approximations to a more subtle quantum universe~\cite{twilight}. 

Note that without something similar to the energy conditions to constrain the spacetimes one wishes to consider as ``physical'', one can construct arbitrarily weird spacetimes containing such exotic objects as warp-drives~\cite{alcubierre, warp-limitations}, traversable wormholes~\cite{mty, visser-wormholes, wormholes-small-violations, roman-wormholes}, singularity-free ``black holes''~\cite{bergmann-roman}, GNACHOs~\cite{gnachos}, violations of the generalized second law~\cite{gsl}, violations of cosmic censorship~\cite{censor}, and even time machines~\cite{thorne-cpc, hawking-cpc, visser-cpc}. Because of the need to keep such oddities somewhat constrained, and avoid a complete free-for-all, much work has gone into developing ``averaged energy conditions'' of various types~\cite{thoughts, q-anec, volume-aec}, though even here there are issues associated with quantum field theory anomalies~\cite{anomalies}.

In particular, within the framework of semi-classical General Relativity (GR), the Quantum Inequalities (QIs) are an extremely important tool for constraining both negative energies,  and more generally exotic phenomena which violate the classical energy conditions. 
Originally introduced by Ford~\cite{Ford:1990id}, and subsequently extensively investigated by Ford and Roman~\cite{Ford+Roman, Ford:1996er}, and their collaborators~\cite{Ford+Roman+collaborators, others}, the QIs impose a lower bound for the integral of the renormalized stress-energy tensor along a suitable time interval, weighted by a suitable test function.  Initially the test function was chosen to be a Lorentzian function~\cite{Ford:1996er}, but subsequently Flanagan found a result independent of the specific choice of the test function in flat (1+1) space-time~\cite{Flanagan:1997gn}. With a distinct approach, Fewster and Eveson also obtained a bound independent of the weight function in 4 and 2 dimensional Minkowski space~\cite{Fewster:1998pu}, though their result in 2 dimensions was somewhat weaker than that of Flanagan~\cite{Flanagan:1997gn}.

A related but distinct restriction for negative energies is imposed by the Quantum Interest Conjecture (QIC), introduced by Ford and Roman~\cite{Ford:1999qv}. Informally speaking, the QIC states that ``overall'' the energy density must be positive, any negative energy density in one region must be overcompensated by positive energy in  another region.
When restricted to isolated pulses of energy, the QIC not only constrains the amount of negative energy in a pulse, but also the time interval between any negative energy pulse and the larger positive energy pulse that \emph{must} also be present. The (positive) nett energy of the two pulses is called the Quantum Interest, and grows monotonically with the time separation. Initially Ford and Roman used delta function energy pulses to prove the conjecture, and it had not been generalized to arbitrarily shaped energy pulses in  4 dimensional flat space-time,  nor was it initially formulated for curved space-time. 

A much more general approach to the QIC that is not restricted to isolated pulses was developed by Fewster and Teo in~\cite{Fewster:1999kr}, wherein they created a Sobolev-space-based technical apparatus to reinterpret the QIs and the QIC as eigenvalue problems for a one-dimensional quantum mechanical pseudo-Hamiltonian.  This technique relates the (1+1) dimensional QIs and QIC to the nonexistence bound states for the Schr\"{o}dinger equation (SDE) in one dimension, while the (3+1) dimensional QIs and QIC are related to the nonexistence bound states for the \emph{bi-harmonic} Schr\"odinger equation (bSDE). Within this framework, Fewster and Teo~\cite{Fewster:1999kr}, and later Teo and Wong~\cite{Teo:2002ne}, were able to prove the QIC in (1+1) dimensional Minkowski space; however they could not extend their proof to the (3+1) dimensional case. In Teo and Wong's approach~\cite{Teo:2002ne} critical use is made of a theorem by Simon~\cite{Simon}, regarding the existence of bound states for the 1 dimensional SDE. That is, Simon's theorem is used to prove a version of the QIC in (1+1) dimensions, but the generalization to (3+1) dimensions is troublesome. 

Furthermore, in~\cite{Fewster:1999kr} Fewster and Teo proved that is not possible for a positive delta function pulse to compensate a negative delta function pulse in 4 dimensions, regardless how small the time separation between them could be.
Additionally, in~\cite{Pretorius:1999nx}  Pretorius  found an upper bound for the separation between two general energy pulses by applying a scaling argument to the test function. With this strategy, Pretorius also showed that the Quantum Interest grows almost linearly as the pulse separation increases, which is the same result obtained in~\cite{Ford:1999qv}.

Turning to the current article: Using the approach explained in~\cite{Fewster:1999kr}, we shall provide a simple variational proof for a variant of the QIC in 4 dimensional flat space-time, (and in fact in any even dimensional spacetime), by proving the equivalent of Simon's theorem for the \emph{bi-harmonic} Schr\"{o}dinger equation (and more generally for the \emph{multi-harmonic} Schr\"odinger equation). We shall do this via a variational argument coupled with a power series expansion of the test function used in the related one-dimensional pseudo-Hamiltonian problem. 

The layout is as follows: In section II we explore the QIs and the QIC from the point of view of eigenvalue problems,  to obtain the related Schr\"odinger equations (SDE and bSDE). Then in section III we will use an appropriate class of  test functions to prove Simon's theorem via variational techniques applied to a power series expansion, and then in section IV repeat this strategy for the (3+1) dimensional related problem to prove the equivalent of Simon's theorem for the bSDE. (Furthermore, as shown in the appendix,  this naturally leads to a version of Simon's theorem for \emph{multi-harmonic} Schr\"odinger equations.) When translated back to the underlying even-dimensional Minkowski spacetime this guarantees the \emph{existence} of quantum interest, though this particular calculation is mute as to the amount of interest and ``interest rate''.  We conclude with a brief discussion in section V.

\section{Quantum Inequalities and the Quantum interest Conjecture} 
\subsection{The Quantum Inequalities}

The Quantum Inequalities give a lower bound for the expectation value of the renormalized stress-energy tensor (in a quantum state $\psi$) when evaluated along a complete timelike geodesic,
\begin{eqnarray}
\label{ec1}
I_{\psi,w} &\equiv&\int_{-\infty}^{\infty}\left<\,T_{00}^{ren}(t,0)\,\right>_{\psi}w(t)\,\d t
\\
&\equiv&\oint\left<\,T_{00}^{ren}(t,0)\,\right>_{\psi}w(t)\,\d t.
\end{eqnarray}
Here $w(t)$ is a more-or-less freely-specifiable test function (weighting function), which is non-negative and integrates to unity:  $\int_{-\infty}^{+\infty} w(x) \; \d x = 1$. Initially, the stress-tensor is treated in the test-field limit, so the background is simply taken to be Minkowski space, and the timelike geodesic is specified by the 4-velocity $V^a=(1;\vec 0)$.  Since our integrals almost always run from $-\infty$ to $+\infty$ we shall for brevity often use the shorthand symbol $\oint = \int_{-\infty}^{+\infty}$.

In their initial analysis, Ford and Roman used a Lorentzian test function to yield to a lower bound which depend on modified Bessel functions~\cite{Ford:1996er}. However a more general inequality was found by Flanagan in (1+1) dimensional flat space-time~\cite{Flanagan:1997gn}, where he used Quantum Field Theory for a  massless scalar field minimally coupled to gravity to find the \emph{optimum} lower bound independent of the specific choice of the test function,
\begin{equation}
\label{Fla}
I_{\psi,w}\geq-\frac{1}{24\pi}\oint\frac{(w'(t))^2}{w(t)}\,\d t.
\end{equation}
Fewster and Eveson used a distinct approach~\cite{Fewster:1998pu} to yield inequalities in (1+1) dimensions,
\begin{equation}
 I_{\psi,w}\geq-\frac{1}{16\pi}\oint\frac{(w'(t))^2}{w(t)}\,\d t,
\end{equation}
and in (3+1) dimensions,
\begin{equation}
\label{Few}
I_{\psi,w}\geq-\frac{1}{16\pi}\oint\left([w^{1/2}]''(t)\right)^2\,\d t.
\end{equation}
It is obvious that the better bound in (1+1) dimensions is given by Flanagan's result rather than the one by Fewster and Eveson. However the most general and optimum bound in (3+1) dimensional flat space-time, for a massless scalar field minimally coupled to gravity, is shown in equation (\ref{Few}).

Combining the results of Flanagan, and Fewster and Eveson, is possible to rewrite the QIs in $2m$-dimensional spacetime in the more general form~\cite{Fewster:1999kr}, (writing $w(t) = f(t)^2$ to automatically enforce the positivity constraint), 
\begin{equation}
\label{QI1}
\oint\left<\,T_{00}^{ren}(t,0)\,\right>_{\psi}\,[f(t)]^2\,\d t
\geq
-\frac{1}{c_m}\oint[D^{m}f(t)]^2\,\d t,
\end{equation}
where $D$ is the derivative operator, and the various different constants are given by
\begin{eqnarray}
 c_m= \left\lbrace \begin{array}{cc}
6\pi &m=1; \\
m\pi^{m-1/2}2^{2m}\Gamma(m-\frac{1}{2})& m\geq 2;
\end{array} \right.
\end{eqnarray}
and where the weighting function is now normalized to $\int_{-\infty}^{+\infty} |f(x)|^2 \; \d x = 1$.
The case $m=1$ is equivalent to equation (\ref{Fla}), while equation (\ref{Few}) is equivalent to  $m=2$. After integrating equation (\ref{QI1}) by parts, the QIs become a statement regarding the non-existence of negative eigenvalues for a  one-dimensional pseudo-Hamiltonian,
\begin{equation}
 \left<f|H|f\right>\geq0,
\end{equation}
where,
\begin{equation}
\label{Ham}
 H=(-1)^mD^{2m}+c_m\,\left<\,T_{00}^{ren}(t,0)\,\right>_{\psi}.
\end{equation}
The quantity $c_m\,\left<\,T_{00}^{ren}(t,0)\,\right>_{\psi}$ can effectively be viewed as an  ``potential'' $V$ for a quantum mechanical system,
\begin{equation}
\label{V}
 V\equiv c_m\,\left<\,T_{00}^{ren}(t,0)\,\right>_{\psi},
\end{equation}
and the QIs become the statement
\begin{equation}
 \left<f|  \left\{  (-1)^mD^{2m}+ V \right\}  |f\right>\geq0,
\end{equation}
We shall now use the framework of quantum mechanics to analyze the QIs and the QIC as an eigenvalue problem.

\subsection{The QIC as an eigenvalue problem}

As adopting this point of view allows us to use the mathematical background of one-dimensional  quantum mechanics, we will change to the standard quantum mechanical notation. For instance,  the differential operator (\ref{Ham}) becomes the quantum pseudo-Hamiltonian
\begin{equation}
\label{H2}
 H= P^{2m}+\,V,
\end{equation}
where $P$ and $V$ are operators on the usual Hilbert space of square-integrable functions. The eigenvalue problem for this Hamiltonian, in coordinates, is the ODE [ordinary differential equation] (we have effectively set $\hbar/2m\rightarrow1$ to simplify the algebra),
\begin{equation}
\label{ODE}
 (-1)^m\,\frac{\d^{2m}}{\d x^{2m}}\,\varphi(x)+V(x)\,\varphi(x)=E\,\varphi(x),
\end{equation}
for the eigenfunctions $\varphi(x)$ which again belong to the usual Hilbert space of square-integrable functions.  (If desired, a more rigorous approach can be followed, as  in~\cite{Fewster:1999kr}). Taking $m=1$, (\ie, (1+1) spacetime dimensions), we recover the one-dimensional time-independent Schr\"odinger equation (SDE)
\begin{equation}
\label{SDE}
 -\frac{\d^2}{\d x^2}\,\varphi(x)+V(x)\,\varphi(x)=E\,\varphi(x),
\end{equation}
while, for $m=2$, (\ie, (3+1) spacetime dimensions), we obtain the one-dimensional  time-independent \emph{bi-harmonic} Schr\"odinger Equation (bSDE),
\begin{equation}
\label{bSDE}
 \frac{\d^4}{\d x^4}\,\varphi(x)+V(x)\,\varphi(x)=E\,\varphi(x).
\end{equation}
For the case of the SDE, there is a theorem by Simon~\cite{Simon} which, by imposing a constraint on the potential, ensures the existence of a bound state for the Hamiltonian.  In a simple form it reads:

\bigskip
\noindent{\bf Simon's Theorem:} \ \\
\emph{
 Let $V(x)$ obey $\int_{-\infty}^{\infty}\,(1+x^2)\,|V(x)|\,dx<\infty$, with $V(x)$ not zero almost everywhere. Then  $H=-\d^2/\d x^2+V(x)$ has a negative eigenvalue if
\begin{equation}
\label{Sim}
 \int_{-\infty}^{\infty}V(x)\,\d x\leq0.
\end{equation}
\hfill $\Box$
}

\bigskip
\noindent
With the help of this theorem, it is possible to generalize the QIC to a wider set of energy pulses, and not only be restricted to $\delta$-function pulses as in~\cite{Ford:1999qv}. 

In order to better understand the QIC in this framework, first we have to point out that a potential which fulfills the QIs must violate condition (\ref{Sim}). 
Now let us rewrite the potential as
\begin{equation}
 \label{V+-}
V(x)=V(x)_{+}-V(x)_{-}.
\end{equation}
This splits the potential into its positive part minus its negative contributions, with $V(x)_{\pm}\geq0$. Then, if the potential \emph{violates} condition (\ref{Sim}), from (\ref{V+-}) we have
\begin{equation}
\label{V>V}
 \oint V(x)_+ \; \d x > \oint V(x)_-   \; \d x .
\end{equation}

This means, if we go back to the GR terminology, that the QIs imply that the expectation value of the renormalized stress-energy tensor must satisfy
\begin{equation}
\oint\left<\,T_{00}^{ren}(t,0)\,\right>_{\psi}\;\d t > 0,
\end{equation}
which is slightly \emph{stronger} than the averaged weak energy condition (AWEC). Splitting the energy density into its positive and negative parts
\begin{equation}
\label{rho+-}
\left<\,T_{00}^{ren}(t,0)\,\right> = 
\left<\,T_{00}^{ren}(t,0)\,\right> _+ - \left<\,T_{00}^{ren}(t,0)\,\right> _-,
\end{equation}
this implies
\begin{equation}
\label{rho>rho}
 \oint\left<\,T_{00}^{ren}(t,0)\,\right>_+ \; \d x > 
 \oint\left<\,T_{00}^{ren}(t,0)\,\right>_-   \; \d x .
\end{equation}
That is, the positive part of a stress tensor that satisfies the QIs, must always \emph{overcompensate} its negative part, this is one version of QIC. (Of course, there are significant elements missing from this form of the conjecture, as compared with the original form and Pretorius'  scaling argument~\cite{Pretorius:1999nx}. Specifically there is no immediate way to extract a bound for the time separation between the energy pulses and the Quantum Interest as a function of this separation. However, a benefit off the current discussion is that we are not limited to delta function pulses of pulses of compact support.)   

In the next section we will prove Simon's theorem for the SDE by expanding an appropriate class of test functions in a power series; this strategy will be used again for the bSDE in section 4, and in the appendix a similar result will be obtained in arbitrary even-dimensional Minkowski spacetime.

\section{1+1 dimensional Minkowski space}

In order to prove Simon's theorem,  avoiding most of the technical issues arising in the formal proof~\cite{Simon}, and keeping in mind the idea of generalizing the theorem to the bSDE, and even higher-derivative ``Hamiltonians'', we shall use a power series expansion of the test function in the SDE to find the conditions for  the potential $V$ to bind.

\subsection{Gaussian wave-function}

Let us start, for simplicity and clarity, with a Gaussian test function,
\begin{equation}
\label{Gau}
 \varphi_{test}=
 \left[\frac{\exp\{{-{(x-\mu)^2}/{(2\sigma^2)}}\}}{\sqrt{2\pi} \sigma}\right]^{1/2},
\end{equation}
which automatically enforces the unit normalization $\oint|\varphi(x)|^2\,\d x=1$.  Therefore, a variational argument implies that for the lowest eigenvalue of the SDE one has:
\begin{equation}
E\leq\oint\left[\varphi'(x)^2+V(x)\,\varphi(x)^2\right]\d x.
\end{equation}
Then the kinetic term is $\oint\varphi'(x)^2=1/\sigma^2$, whereas (assuming all the moments $\left|\oint x^{2n}\,V(x)\,dx\right|<\infty$ so that the expansion makes sense)
\begin{eqnarray}
&&\oint V(x)\,\varphi(x)^2\d x=
\nonumber\\
&&
\quad 
\frac{1}{\sqrt{2\pi}\,\sigma}\sum_{n=0}^{\infty}\frac{(-1)^n}{n!(2\sigma^2)^n}\times\oint V(x)\,(x-\mu)^{2n}\d x.\qquad
\end{eqnarray}
Interchanging the integral and the summation is justified since the Gaussian function $\varphi(x)^2$ is an analytic function of $x$ whose Taylor series has an infinite radius of convergence.

For a fixed $\mu$ we now see that:
\begin{equation}
 E\,\sigma^2\leq1+\frac{\sigma}{\sqrt{2\pi}}\oint V(x)\,\d x+O(1/\sigma).
\end{equation}
From this last expression, we can see that if 
\begin{equation}
 \oint V(x)\,\d x<0,\
\end{equation}
then for $\sigma$ sufficiently large we can guarantee $E\,\sigma^2<0$, (hence $E<0$ and the potential will bind, and so violate the QIs).  This is close to Simon's condition. (We currently have a $<$, rather than Simon's $\leq$. The only ``difficult'' thing about Simon's theorem is proving the existence of a bound state in the marginal $=$ case. Note that the weaker result we have here is already enough, in GR language, to assure that the stress tensor satisfies the AWEC.)

\subsection{Generic class of test functions}

To deal with the marginal case, we proceed by using the fact that we are free to  \emph{choose} the test function in a more-or-less arbitrary manner, and so to use this freedom obtain tighter constraints. Let us start by picking some function $g(x)$ that is analytic on the entire real line (infinite radius of convergence) and then constructing the piecewise analytic function
\begin{equation}
h(x) = g(\, |x|\, ).
\end{equation}
Note that $h(x)$ need not, and typically will not, be analytic at zero.
We then enforce the normalization
\begin{equation}
\label{norm}
 \oint h(x)\,\d x=1,\quad\oint x\;h(x)\, \d x=0,\quad\oint x^2\,h(x)\, \d x=1.
\end{equation}
Suitable examples of such functions are 
\begin{equation}
h(x) =  {\exp(-\sqrt{2}\,|x|\,)\over\sqrt{2}},
\end{equation}
and
\begin{equation}
h(x) = \sum_{i=1}^N c_i \; \exp(-|x|/d_i),
\end{equation}
subject to 
\begin{equation}
2 \sum_{i=1}^N c_i \,d_i = 1; \qquad 4 \sum_{i=1}^N c_i \,d_i^3 = 1. 
\end{equation}
We now choose
\begin{equation}
\label{Gen}
 \varphi_{test}=\sqrt{\frac{h\left({[x-\mu]}/{\sigma}\right)}{\sigma}} = 
 \sqrt{\frac{g\left({|x-\mu|}/{\sigma}\right)}{\sigma}},
\end{equation}
whence
\begin{equation}
 \oint|\varphi(x)|^2\,\d x=1,\quad\oint x\;|\varphi(x)|^2\,\d x=\mu,
\end{equation}
and
\begin{equation} 
\oint(x-\mu)^2\;|\varphi(x)|^2\;\d x=\sigma^2.
\end{equation}
Moreover, $\oint\varphi'(x)^2dx=\kappa/\sigma^2$, with
\begin{equation}
\label{kap1}
 \kappa\equiv\frac{1}{4}\oint\frac{h'(x)^2}{h(x)} \; \d x.
\end{equation}
The numeric value of $\kappa$ depends on the choice of the test function and further optimizations may be useful. For now, let us focus on the Taylor expansion of $g(x)$:
\begin{equation}
\label{gexp0}
 g(x)=\sum_{n=0}^\infty a_n\,x^n.
\end{equation}
This is assumed to exist (by analyticity), and to converge on the entire real line. We now deduce the existence of a similar power series expansion for $h(x)$:
\begin{equation}
\label{gexp}
 h(x)= g(\,|x|\,) = \sum_{n=0}^\infty a_n\,|x|^n.
\end{equation}
This is now an expansion in the variable $|x|$. Since it depends on $|x|$, this is \emph{not} a Taylor series for $h(x)$, but by construction it is convergent over the entire real line.  Then for the shifted and rescaled test function we have 
\begin{equation}
 |\varphi_{test}|^2=\frac{1}{\sigma}\sum_{n=0}^\infty a_n\left[\frac{|x-\mu|}{\sigma}\right]^n.
\end{equation}
Assuming all the moments $|\oint |x|^n\,V(x)\,\d x|<\infty$, so that the expansion makes sense, we furthermore have
\begin{equation}
 E\leq\frac{\kappa}{\sigma^2}+
 \frac{1}{\sigma}\sum_{n=0}^\infty\frac{a_n}{\sigma^n}\oint V(x)\;|x-\mu|^n\; \d x.
\end{equation}
In interchanging the summation and the limit we have first split $\oint \equiv \int_{-\infty}^{+\infty} =  \int_{-\infty}^\mu  + \int_\mu^{+\infty}$, and then appealed to the analyticity of $g(x)$ and $g(-x)$ on the appropriate ranges, finally recombining the two sets of integrals to run over the entire real line.

We now have the potential for additional relevant terms to appear in the series expansion for the variational bound on the lowest eigenvalue $E$, and is convenient to place some restrictions on the coefficients $a_n$ appearing in the series expansion. 

First, if $a_0=0$, then 
\begin{equation}
h(x)\sim a_1\,|x|+\dots,
\end{equation}
so that
\begin{equation}
{[h'(x)]^2\over h(x)} \sim {a_1\over |x|} +\dots.
\end{equation}
Thus  in order for the integral defining $\kappa$ [equation (\ref{kap1})] to converge at $x=0$ it is necessary to chose  $a_1=0$. So in fact  
\begin{equation}
h(x)\sim a_2\,x^2+\dots,
\end{equation}
and
\begin{equation}
{[h'(x)]^2\over h(x)} \sim 4 \, a_2 +\dots.
\end{equation}
 In this situation we then cannot extract much information, since the SDE now merely gives
\begin{equation}
 E\,\sigma^2\leq\kappa+O(\sigma^{-1}).
\end{equation}
Hence, to extract useful information, we need $a_0\neq0$, and so (since probability densities are always positive), $a_0>0$.

In contrast, for $a_0>0$, the integral defining $\kappa$ [equation (\ref{kap1})] always converges at $x=0$, since 
\begin{equation}
h(x)\sim a_0+a_1\,|x| + \cdots,
\end{equation}
and so 
\begin{equation}
{[h'(x)]^2\over h(x)} \sim {a_1^2\over a_0} +\dots.
\end{equation}
Thus we have
\begin{equation}
 E\,\sigma^2\leq\kappa+\sigma\,a_0\oint V(x)\,\d x+a_1\oint|x-\mu|\,V(x)\,\d x
 +O(1/\sigma).
\end{equation}
Therefore, with $\sigma$ sufficiently large the only important contributions come from the first two terms ($n=0$, and $n=1$) of the power series expansion.

Here, as for the Gaussian test functions, the condition $\oint V(x)\,\d x<0$ implies binding; which means that the expectation value of the  renormalized stress-energy tensor must fulfill the AWEC. However, if we consider the borderline of AWEC violation, \ie, $\oint V(x)\,\d x=0$, then the next term in the expansion is now no longer neglectable:
\begin{equation}
 E\,\sigma^2\leq\kappa+a_1\oint|x-\mu|\,V(x)\,\d x
 +O(1/\sigma).
\end{equation}
But recall that completely $a_1$ is arbitrary,  both in \emph{sign} and in \emph{magnitude}. That is, if for \emph{any} value of $\mu$
\begin{equation}
 \oint |x-\mu|\,V(x)\,\d x\neq0,
\end{equation}
then this implies the existence of a bound state for the SDE. This is a very strong constraint on the potential and it quickly yields a proof of Simon's theorem. Taking the converse of the above, if $\oint V(x) \; \d x=0$ then to \emph{prevent} the occurrence of a bound state we must have
\begin{equation}
\forall \mu: \qquad  \oint |x-\mu|\,V(x)\,\d x=0.
\end{equation}
Differentiating twice with respect to $\mu$ we get
\begin{eqnarray}
\forall \mu:\qquad  2\oint\delta(x-\mu)\,V(x)\,dx &=&0\nonumber\\
\Rightarrow V(\mu)&=&0.
\end{eqnarray}
Thus any non-zero potential binds if its integral is null.  Combining this with the fact that $\oint V(x)\,dx<0$ also implies binding, now provides  a simple variational proof of Simon's theorem, equation (\ref{Sim}), though under the much stronger technical conditions that all the moments exist $|\oint |x|^n \, V(x)\; \d x | < \infty$. It is these stronger technical conditions --- which are still physically quite reasonable and are certainly satisfied by isolated pulses of stress energy --- that will make it easy for us to extend Simon's theorem to the \emph{bi-harmonic} Schr\"odinger equation, and so lead to a (3+1) dimensional version of the QIC.

\section{3+1 dimensional Minkowski space}

To generalize the condition given by Simon's theorem to the bSDE, we proceed as before, expanding appropriate  test functions in power series. Applying a variational argument to the bSDE we find that the lowest eigenvalue satisfies
\begin{equation}
\label{E4}
 E\leq\oint\left[\,\varphi''(x)^2+V(x)\;\varphi(x)^2\,\right]\d x,
\end{equation}
assuming that the test functions are normalized. We now use this relation to probe for  the existence of a bound state for the pseudo-Hamiltonian for the bSDE, which is ultimately related to the 4 dimensional flat space-time QIs and the QIC.
\subsection{Gaussian wave-function}

We start with the normalized Gaussian test function previously used for the SDE,  equation (\ref{Gau}).  The kinematic term yields $\oint\varphi''(x)^2\d x=\kappa^2/\sigma^4$, with $\kappa$ a numeric constant. Assuming all appropriate moments exist, we again expand  the test function in a power series, whence
\begin{eqnarray}
&&\oint V(x)\;\varphi^2(x)\,dx=
\nonumber\\
&&\quad
\frac{1}{\sqrt{2\pi}\, \sigma}
\sum_{n=0}^{\infty}\frac{(-1)^n}{n!(2\sigma^2)^n}\times\oint V(x)\;(x-\mu)^{2n}\d x.
\qquad
\end{eqnarray}
Then, from (\ref{E4})
\begin{eqnarray}
E\,\sigma^4&\leq&\kappa^2+\frac{\sigma^3}{\sqrt{2\pi}}\oint V(x)\,\d x
\nonumber\\
&&
-\frac{\sigma}{2\sqrt{2\pi}}\oint V(x)\;(x-\mu)^2\,\d x
 +O(1/\sigma).\qquad
\end{eqnarray}
As for the SDE, in this case also $\oint V(x)\,\d x<0$ implies that the potential binds for a sufficiently large $\sigma$. This gives us most of Simon's theorem, apart from the extremal case where the integral vanishes.

However, even with the Gaussian test function, in this bSDE case we recover more information than for the SDE. Specifically if $\oint V(x)\,\d x=0$, then the next term of the expansion is important, and to guarantee the \emph{absence} of a bound state we must also enforce:
\begin{equation}
\forall \mu: \qquad  \oint(x-\mu)^2\; V(x) \; \d x\leq0.
\end{equation}
This constraint, while certainly significant, is not quite strong enough to imply Simon's theorem. 

\subsection{Generic class of test functions}

Finally, for the bSDE, we consider the same normalization,  (\ref{norm}), and the same choice for the generic test function, as we used for the SDE, (\ref{Gen}). Note
\begin{equation}
 \oint \varphi''(x)^2\;\d x=\frac{\kappa}{\sigma^4},
\end{equation}
with
\begin{eqnarray}
\label{kap2}
 \kappa&\equiv&
 \oint \left\{ \left[\sqrt{h}\right]''(x)\right\}^2 \; \d x
 \\
 &=&\frac{1}{16}\oint\frac{\left[2h(x)\,h''(x)-h'(x)^2\right]^2}{h(x)^3}\,\d x.
\end{eqnarray}
In view of the fact that we have chosen $h(x) = g(\,|x|\,)$ this becomes (note the delta function arising from twice differentiating the absolute value) 
\begin{eqnarray}
\label{kap3}
&& 
\!\!\!\!\!
\kappa\equiv\frac{1}{16}\times 
\\
&&
\!\!\!\!\!
\oint\frac{\left\{2g(|x|)\,\left[g''(|x|)+g'(|x|)\,\delta(x)\right]-g'(|x|)^2\right\}^2}{g(|x|)^3}\,\d x.
\nonumber
\end{eqnarray}
In order to derive a useful bound we will want $\kappa$ to be finite, which means that we want the coefficient of the delta function to be zero, that is $a_1=0$.  To make the integral converge at zero we also want $a_0\neq0$, and in fact must have $a_0>0$. In contrast, $a_2$ and  $a_3$ are unconstrained as to sign and magnitude and we have
\begin{eqnarray}
\label{Eexp}
 E\,\sigma^4&\leq&\kappa+a_0\;\sigma^3\oint V(x)\,\d x\nonumber\\
&+&a_2\;\sigma\oint V(x)\;|x-\mu|^2\d x\nonumber\\
&+&a_3\;\oint V(x)\;|x-\mu|^3\d x+O(1/\sigma).
\end{eqnarray}
As before, by suitably choosing a sufficiently large $\sigma$, we see that $\oint V(x)\,\d x<0$ implies binding.

To extract further information from (\ref{Eexp}) it is necessary to check what happens if we set $\oint V(x)\,\d x=0$. Thereafter the next two terms in the expansion become relevant, however both the \emph{sign} and \emph{magnitude} of $a_2$ and $a_3$ are freely specifiable, hence (for any $\mu$) either
\begin{equation}
 \oint |x-\mu|^2\; V(x)\;\d x\neq0,
\end{equation}
or
\begin{equation}
 \oint |x-\mu|^3\; V(x)\;\d x\neq0,
\end{equation}
is a \emph{sufficient} condition for the Hamiltonian to \emph{have} a bound state. 

Conversely, if $\oint V(x)\; \d x =0$, a \emph{necessary} condition for the \emph{absence} of a bound state is that  
\begin{equation}
\label{E:V2}
\forall \mu: \qquad  \oint |x-\mu|^2\; V(x)\;\d x=0,
\end{equation}
and
\begin{equation}
\forall \mu: \qquad  \oint |x-\mu|^3\; V(x)\;\d x=0.
\end{equation}
Now if we differentiate the last expression with respect to $\mu$ then
\begin{equation}
\forall \mu: \qquad  \oint |x-\mu|^2  \; \mathrm{sign}(x-\mu) \; V(x)\;\d x=0,
\end{equation}
whence, combining with Eq. (\ref{E:V2}), we have
\begin{equation}
\forall \mu: \qquad  \int_\mu^\infty |x-\mu|^2  \; V(x)\;\d x=0.
\end{equation}
If we repeatedly differentiate the last expression with respect to $\mu$ then
\begin{equation}
\forall \mu: \qquad  \int_\mu^\infty |x-\mu|  \; V(x)\;\d x=0.
\end{equation}
\begin{equation}
\forall \mu: \qquad  \int_\mu^\infty  \; V(x)\;\d x=0.
\end{equation}
and finally $V(x) =0$.
That is,  if $\oint V(x)\; \d x =0$, a \emph{necessary} condition for the \emph{absence} of a bound state is that   $V(x) =0$. Conversely if  $\oint V(x)\; \d x =0$ and  $V(x)\not\equiv0$, then this is a \emph{sufficient} condition for the \emph{presence} of a bound state.
This proves the equivalent of Simon's theorem for the 4th-order bSDE and, furthermore proves the QIC in (3+1) dimensional Minkowski space. 

Note that the version of the QIC that we have proved is this: \qquad The QI's imply that \emph{either} $\left<\,T_{00}^{ren}(t,0)\,\right>\equiv0$ everywhere along the world-line, \emph{or}
\begin{equation}
\label{E:qic}
\oint\left<\,T_{00}^{ren}(t,0)\,\right>_{\psi}\;\d t > 0,
\end{equation}
which we emphasize is slightly \emph{stronger} than the AWEC. Splitting the energy density into its positive and negative parts, this implies that as long as $\left<\,T_{00}^{ren}(t,0)\,\right>$ is not identically zero along the world line, then
\begin{equation}
\label{rho>rho2}
 \oint\left<\,T_{00}^{ren}(t,0)\,\right>_+ \; \d x > 
 \oint\left<\,T_{00}^{ren}(t,0)\,\right>_-   \; \d x .
\end{equation}
That is, any negative energy ``loan'' is overcompensated elsewhere along the world line.

\section{Discussion}

With the variant of Simon's theorem that we have now proved for the bSDE, it is possible  to reformulate the QIC for a more general range of energy pulses in (3+1) dimensional flat space-time, as has been already done for the (1+1) dimensional case. In flat space-time, an energy pulse which satisfies the QIs (and so the version of the  QIC discussed above), must also fulfill an AWEC-like inequality. That is, the expectation value of the renormailzed stress-energy tensor shall violate the condition provided by equation (\ref{Sim}), and so satisfy (\ref{E:qic}).

Moreover, using the argument derived from equation (\ref{rho+-}), the QIC can be extended to (3+1) dimensional Minkowski space. Furthermore, from (\ref{rho>rho}) and (\ref{rho>rho2}), we know that the positive contributions of the expectation value of the renormalized stress-energy tensor must overcompesate its negative parts.  The key trade-off in the current argument is that while we have been able to deduce a general result for arbitrary even dimensional Minkowski space, and while we are not limited to delta-function pulses or pulses of compact support, we have on the other hand lost some of the precision information that can be deduced when stronger assumptions are made regarding the temporal distribution of the stress energy.

Furthermore, this proof of the QIC that we have given in flat four dimensional space-time gives only a partial picture of the nature of negative energies and the constraints that can be placed on them, as it does not (yet) include the effects of curved spacetime.  Indeed, for many technical reasons it would be preferable to work with the null energy condition (NEC), rather than the weak energy condition~\cite{q-anec}.
That is: A truly complete formulation of the QIs and QIC should really include curved spacetimes, at the very least (3+1) dimensional curved spacetimes, and work with some version of the null energy condition --- this is the key arena wherein the possibility of exotic phenomena such as  traversable wormholes, warp-drives, and time machines are related to the existence of negative energies and the constraints thereupon.

\section*{Acknowledgments}
This research was supported by the Marsden Fund administered by the Royal Society of New Zealand. GA was additionally supported by a Victoria University of Wellington PhD scholarship.

\appendix
\section{Higher dimensions}
Having now seen the argument in (1+1) and (3+1) dimensions it is clear how to generalize to any even number of dimensions. Consider a $2m$-dimensional (that is, ($[2m-1]+1$) dimensional) spacetime. Then the QIs are equivalent to the statement:
\begin{equation}
\not\!\exists \hbox{ bound state}:  \qquad H = (-1)^m {\d^{2m}\over \d x^{2m}} + V(x).
\end{equation}
Picking the generic class of test functions discussed previously, and applying a variational argument, the ground state energy is bounded by
\begin{equation}
E \leq \oint \left\{ [\varphi^{(m)}(x)]^2 + V(x) \,[ \varphi(x)]^2  \right\}\; \d x.
\end{equation}
The kinetic term is
\begin{equation}
\oint [\varphi^{(m)}(x)]^2  \;  \d x
 = {\kappa^2\over\sigma^{2m}} 
\end{equation}
(where the precise numerical value of $\kappa$ is not important as long as it is finite), whereas (assuming all appropriate moments exist)
\begin{equation}
\oint V \; [\varphi^2]  \; \d x = {1\over\sigma} 
\sum_{n=0}^\infty {a_n\over\sigma^n} \times \oint  V(x)  |x-\mu|^{n} \d x.
\end{equation}
The interchange of the summation and integration is again justified by the assumed piecewise analytic nature of the test function.
Therefore
\begin{eqnarray}
E \,\sigma^{2m} &\leq& \kappa^2 + 
\sigma^{2m-1} \sum_{n=0}^{2m-1} {a_n\over\sigma^n} \times \oint  V(x)  |x-\mu|^{n} \d x
\nonumber
\\
&&
+ O(1/\sigma),
\end{eqnarray}
where there will be some set of constraints on the coefficients $a_n$ to keep $\kappa$ finite.

Now since  $\varphi(x)$ is a test function we are always free to choose $a_0\neq0$, and since probability densities are always positive, this forces $a_0>0$. We are also free to (temporarily) choose all the odd $a_{2n+1}=0$, which is a convenience (since it implies the absence of squared delta functions coming from differentiations of the absolute value function) to guarantee $\kappa$ finite.  Then
\begin{eqnarray}
E \,\sigma^{2m} &\leq& \kappa^2 + 
\sigma^{2m-1} \sum_{n=0}^{m-1} {a_{2n}\over\sigma^{2n}} \times \oint  V(x)  |x-\mu|^{2n} \d x
\nonumber
\\
&&
+ O(1/\sigma),
\end{eqnarray}

But then if $\oint V(x)\, \d x <0$, it follows that for $\sigma$ sufficiently large we can guarantee $E\,\sigma^{2m}<0$, whence $E<0$, and the potential will bind, (thus violating the QIs). This is the easy bit. 
\begin{itemize}
\item In particular, $\not\!\exists$ bound state $\implies  \oint V(x)\, \d x  \geq 0$.
\end{itemize}
Now consider the borderline case  $\oint V(x)\, \d x = 0$. 
We are again free to choose $a_0\neq0$ (and so $a_0>0$), and are \emph{also} free to choose both the \emph{sign} and \emph{magnitude} of all the $a_{2n}$. But this implies the potential will bind unless
\begin{equation}
\forall \mu: \qquad \oint  V(x)  \; |x-\mu|^{2n} \; \d x= 0;  \qquad \forall n \in (0, m-1).
\end{equation}
In particular
\begin{equation}
\forall \mu: \qquad \oint  V(x)  \; |x-\mu|^{2m-2} \; \d x= 0. 
\end{equation}

Now ``turn on'' one of the odd $a_{2n+1}$. Specifically, consider $a_{2m-1}$. Since for $m\geq 2$ we have 
\begin{equation}
{\d^m\over\d x^m} ( |x|^{2m-1} )= 0,
\end{equation}
we can switch on this  $a_{2m-1}$ coefficient without risk of developing a squared delta function in the evaluation of $\kappa$, and so keep $\kappa$ finite. But then since  $a_{2m-1}$ is arbitrary as to sign and magnitude, to prevent binding we must have
\begin{equation}
\forall \mu: \qquad \oint  V(x)  \; |x-\mu|^{2m-1} \; \d x= 0. 
\end{equation}
Differentiating with respect to $\mu$
\begin{equation}
\forall \mu: \qquad \oint  V(x)  \; |x-\mu|^{2m-2} \; \mathrm{sign}(x-\mu) \; \d x= 0,
\end{equation}
whence, combining the two preceeding equations, we see
\begin{equation}
\forall \mu: \qquad \int_\mu^\infty  V(x)  \; |x-\mu|^{2m-2} \; \d x= 0.
\end{equation}
Repeated differentiations with respect to $\mu$ will now eventually yield $V(x)=0$. 

That is,  if  $\oint V(x)\, \d x = 0$ then to prevent a bound state we must have $V(x)=0$.  Conversely, if $\oint V(x)\, \d x = 0$ and $V(x)\not\equiv0$, then there will be a bound state.

\begin{itemize}
\item In particular, $\not\!\exists$ bound state $\implies$ \\
either $V(x) \equiv 0$ or  $\oint V(x)\, \d x  > 0$.\\
\end{itemize}
This now is a version of Simon's theorem appropriate to the \emph{multi-harmonic} SDE, leading to a general even-dimensional version of the QIC:  \emph{Either} $\left<\,T_{00}^{ren}(t,0)\,\right>\equiv0$ everywhere along the world-line, \emph{or}
\begin{equation}
\label{E:qic2}
\oint\left<\,T_{00}^{ren}(t,0)\,\right>_{\psi}\;\d t > 0,
\end{equation}
this last inequality being slightly stronger than the AWEC.




\begin{thebibliography}{90}

 

\bibitem{epstein-glaser-jaffe}
H. Epstein, V. Glaser, and A. Jaffe, 
``Nonpositivity of the energy density in quantized field theories'', 
Nuovo Cimento 36, 1016 (1965).

\bibitem{roman-wec}
T.~A.~Roman,
  ``Quantum Stress Energy Tensors And The Weak Energy Condition,''
  Phys.\ Rev.\  D {\bf 33} (1986) 3526.



\bibitem{twilight}
C.~Barcelo and M.~Visser,
  ``Twilight for the energy conditions?,''
  Int.\ J.\ Mod.\ Phys.\  D {\bf 11} (2002) 1553
  [arXiv:gr-qc/0205066].
  
\bibitem{alcubierre}
 M.~Alcubierre,
  ``The warp drive: hyper-fast travel within general relativity,''
  Class.\ Quant.\ Grav.\  {\bf 11} (1994) L73
  [arXiv:gr-qc/0009013].
  
 \bibitem{warp-limitations}
 M.~J.~Pfenning and L.~H.~Ford,
  ``The unphysical nature of ``warp drive'',''
  Class.\ Quant.\ Grav.\  {\bf 14} (1997) 1743
  [arXiv:gr-qc/9702026].
 \\
  A.~E.~Everett and T.~A.~Roman,
  ``A superluminal subway: The Krasnikov tube,''
  Phys.\ Rev.\  D {\bf 56} (1997) 2100
  [arXiv:gr-qc/9702049].
\\
 F.~S.~N.~Lobo and M.~Visser,
  ``Fundamental limitations on 'warp drive' spacetimes,''
  Class.\ Quant.\ Grav.\  {\bf 21}, 5871 (2004)
  [arXiv:gr-qc/0406083].
  \\
  F.~S.~N.~Lobo and M.~Visser,
  ``Linearized warp drive and the energy conditions,''
  arXiv:gr-qc/0412065.
  
\bibitem{mty}
M.~S.~Morris and K.~S.~Thorne,
 ``Wormholes in space-time and their use for interstellar travel: A tool for
 teaching general relativity,''
  Am.\ J.\ Phys.\  {\bf 56}, 395 (1988).
 \\
M.~S.~Morris, K.~S.~Thorne and U.~Yurtsever,
  ``Wormholes, Time Machines, and the Weak Energy Condition,''
  Phys.\ Rev.\ Lett.\  {\bf 61}, 1446 (1988).
  
\bibitem{visser-wormholes}
M.~Visser,
``Traversable wormholes from surgically modified Schwarzschild spacetimes'',
  Nucl.\ Phys.\  B {\bf 328}, 203 (1989).
  \\
M.~Visser,
  ``Traversable wormholes: Some simple examples'',
  Phys.\ Rev.\  D {\bf 39}, 3182 (1989).
  \\
M.~Visser, \emph{Traversable wormholes: From Einstein to Hawking},   
(AIP Press [now Springer--Verlag], Reading, 1995).

\bibitem{wormholes-small-violations}
M.~Visser, S.~Kar and N.~Dadhich,
  ``Traversable wormholes with arbitrarily small energy condition violations,''
  Phys.\ Rev.\ Lett.\  {\bf 90} (2003) 201102
  [arXiv:gr-qc/0301003].
  \\
  S.~Kar, N.~Dadhich and M.~Visser,
  ``Quantifying energy condition violations in traversable wormholes,''
  Pramana {\bf 63} (2004) 859
  [arXiv:gr-qc/0405103].
  \\
C.~J.~Fewster and T.~A.~Roman,
  ``On wormholes with arbitrarily small quantities of exotic matter,''
  Phys.\ Rev.\  D {\bf 72} (2005) 044023
  [arXiv:gr-qc/0507013].
  \\
  C.~J.~Fewster and T.~A.~Roman,
  ``Problems with wormholes which involve arbitrarily small amounts of exotic
  matter,''
  arXiv:gr-qc/0510079.
  
\bibitem{roman-wormholes}
  T.~A.~Roman,
  ``Inflating Lorentzian wormholes,''
  Phys.\ Rev.\  D {\bf 47} (1993) 1370
  [arXiv:gr-qc/9211012].

\bibitem{bergmann-roman}
T.~A.~Roman and P.~G.~Bergmann,
  ``Stellar Collapse Without Singularities?,''
  Phys.\ Rev.\  D {\bf 28}, 1265 (1983).
  
 \bibitem{gnachos}
 J.~G.~Cramer, R.~L.~Forward, M.~S.~Morris, M.~Visser, G.~Benford and G.~A.~Landis,
  ``Natural wormholes as gravitational lenses,''
  Phys.\ Rev.\  D {\bf 51} (1995) 3117
  [arXiv:astro-ph/9409051].

\bibitem{gsl}
L.~H.~Ford and T.~A.~Roman,
  ``'Cosmic flashing' in four-dimensions,''
  Phys.\ Rev.\  D {\bf 46} (1992) 1328.
  
\bibitem{censor}  
 L.~H.~Ford and T.~A.~Roman,
  ``Moving mirrors, black holes, and cosmic censorship'',
  Phys.\ Rev.\  D {\bf 41} (1990) 3662.

\bibitem{thorne-cpc}
J.~Friedman, M.~S.~Morris, I.~D.~Novikov, F.~Echeverria, G.~Klinkhammer, K.~S.~Thorne and U.~Yurtsever,
  ``Cauchy Problem In Space-Times With Closed Timelike Curves,''
  Phys.\ Rev.\  D {\bf 42}, 1915 (1990).

\bibitem{hawking-cpc}
S.~W.~Hawking,
  ``The Chronology Protection Conjecture,''
  Phys.\ Rev.\  D {\bf 46}, 603 (1992).
    
\bibitem{visser-cpc}
M.~Visser,
  ``The quantum physics of chronology protection,''
 In \emph{The future of theoretical physics and cosmology}, edited by G.~W.~Gibbons,  E.~P.~S.~Shellard, and S.~W.~Hawking, (Cambridge University Press, Cambridge, 2003) , pp. 161--176 [arXiv: gr-qc/0204022].
  \\
  M.~Visser,
  ``The reliability horizon for semi-classical quantum gravity: Metric
  fluctuations are often more important than back-reaction,''
  Phys.\ Lett.\  B {\bf 415}, 8 (1997)
  [arXiv:gr-qc/9702041].
 \\
 M.~Visser,
  ``From wormhole to time machine: Comments on Hawking's chronology protection
  conjecture,''
  Phys.\ Rev.\  D {\bf 47}, 554 (1993)
  [arXiv:hep-th/9202090].
  \\
  M.~Visser,
  ``Hawking's chronology protection conjecture: Singularity structure of the
  quantum stress energy tensor,''
  Nucl.\ Phys.\  B {\bf 416}, 895 (1994)
  [arXiv:hep-th/9303023].
  
  \bibitem{thoughts}
  T.~A.~Roman,
  ``Some thoughts on energy conditions and wormholes,''
  arXiv:gr-qc/0409090.

  \bibitem{q-anec}
  E.~E.~Flanagan and R.~M.~Wald,
  ``Does backreaction enforce the averaged null energy condition in
  semiclassical gravity?,''
  Phys.\ Rev.\  D {\bf 54} (1996) 6233
  [arXiv:gr-qc/9602052].
  \\
  C.~J.~Fewster and T.~A.~Roman,
  ``Null energy conditions in quantum field theory,''
  Phys.\ Rev.\  D {\bf 67} (2003) 044003
  [arXiv:gr-qc/0209036].
  \\
  C.~J.~Fewster, K.~D.~Olum and M.~J.~Pfenning,
  ``Averaged null energy condition in spacetimes with boundaries,''
  Phys.\ Rev.\  D {\bf 75} (2007) 025007
  [arXiv:gr-qc/0609007].
   
  \bibitem{volume-aec}
  L.~H.~Ford, A.~D.~Helfer and T.~A.~Roman,
  ``Spatially averaged quantum inequalities do not exist in four-dimensional
  spacetime,''
  Phys.\ Rev.\  D {\bf 66} (2002) 124012
  [arXiv:gr-qc/0208045].
  
 
\bibitem{anomalies}
M.~Visser,
  ``Scale anomalies imply violation of the averaged null energy condition,''
  Phys.\ Lett.\  B {\bf 349} (1995) 443
  [arXiv:gr-qc/9409043].
   
 

\bibitem{Ford:1990id}
  L.~H.~Ford,
  ``Constraints on negative energy fluxes,''
  Phys.\ Rev.\  D {\bf 43}, 3972 (1991).
  
  \bibitem{Ford+Roman}
  L.~H.~Ford and T.~A.~Roman,
  ``Averaged Energy Conditions And Quantum Inequalities,''
  Phys.\ Rev.\  D {\bf 51} (1995) 4277
  [arXiv:gr-qc/9410043].
  \\
  L.~H.~Ford and T.~A.~Roman,
  ``Quantum Field Theory Constrains Traversable Wormhole Geometries,''
  Phys.\ Rev.\  D {\bf 53} (1996) 5496
  [arXiv:gr-qc/9510071].
  \\
  L.~H.~Ford and T.~A.~Roman,
  ``Motion of inertial observers through negative energy,''
  Phys.\ Rev.\  D {\bf 48} (1993) 776
  [arXiv:gr-qc/9303038].
  

\bibitem{Ford:1996er}
  L.~H.~Ford and T.~A.~Roman,
  ``Restrictions on negative energy density in flat spacetime,''
  Phys.\ Rev.\  D {\bf 55}, 2082 (1997)
  [arXiv:gr-qc/9607003].

\bibitem{Ford+Roman+collaborators}
L.~H.~Ford, M.~J.~Pfenning and T.~A.~Roman,
  ``Quantum inequalities and singular negative energy densities,''
  Phys.\ Rev.\  D {\bf 57} (1998) 4839
  [arXiv:gr-qc/9711030].
 \\   
  A.~Borde, L.~H.~Ford and T.~A.~Roman,
  ``Constraints on spatial distributions of negative energy,''
  Phys.\ Rev.\  D {\bf 65} (2002) 084002
  [arXiv:gr-qc/0109061].
  
\bibitem{others}
C.~J.~Fewster and L.~W.~Osterbrink,
  ``Quantum Energy Inequalities for the Non-Minimally Coupled Scalar Field,''
  J.\ Phys.\ A  {\bf 41} (2008) 025402
  [arXiv:0708.2450 [gr-qc]].
 \\ 
  C.~J.~Fewster,
  ``Quantum energy inequalities in two dimensions,''
  Phys.\ Rev.\  D {\bf 70} (2004) 127501
  [arXiv:gr-qc/0411114].
\\
E.~E.~Flanagan,
  ``Quantum inequalities in two dimensional curved spacetimes,''
  Phys.\ Rev.\  D {\bf 66} (2002) 104007
  [arXiv:gr-qc/0208066].
 
\bibitem{Flanagan:1997gn}
  E.~E.~Flanagan,
  ``Quantum inequalities in two dimensional Minkowski spacetime,''
  Phys.\ Rev.\  D {\bf 56}, 4922 (1997)
  [arXiv:gr-qc/9706006].

\bibitem{Fewster:1998pu}
  C.~J.~Fewster and S.~P.~Eveson,
  ``Bounds on negative energy densities in flat spacetime,''
  Phys.\ Rev.\  D {\bf 58}, 084010 (1998)
  [arXiv:gr-qc/9805024].

\bibitem{Ford:1999qv}
  L.~H.~Ford and T.~A.~Roman,
  ``The quantum interest conjecture,''
  Phys.\ Rev.\  D {\bf 60}, 104018 (1999)
  [arXiv:gr-qc/9901074].

\bibitem{Fewster:1999kr}
  C.~J.~Fewster and E.~Teo,
  ``Quantum inequalities and `quantum interest' as eigenvalue problems,''
  Phys.\ Rev.\  D {\bf 61}, 084012 (2000)
  [arXiv:gr-qc/9908073].

\bibitem{Teo:2002ne}
  E.~Teo and K.~F.~Wong,
  ``Quantum interest in two dimensions,''
  Phys.\ Rev.\  D {\bf 66}, 064007 (2002)
  [arXiv:gr-qc/0206066].

\bibitem{Simon}
B.~Simon, 
``The bound states of weakly coupled Schr\"odinger operators in one and two dimensions'', 
Annals Phys.\  {\bf 97}, 279 (1976).

\bibitem{Pretorius:1999nx}
  F.~Pretorius,
  ``Quantum interest for scalar fields in Minkowski spacetime,''
  Phys.\ Rev.\  D {\bf 61}, 064005 (2000)
  [arXiv:gr-qc/9903055].
  
  
 
 
   
    
    
  
    
  
\end{thebibliography}
\end{document}